RESEARCH ARTICLE

WILEY

# Tree-ring stable isotopes and radiocarbon reveal pre- and post-eruption effects of volcanic processes on trees on Mt. Etna (Sicily, Italy)

Ruedi Seiler[1,2] | Irka Hajdas[3] | Matthias Saurer[1,4] | Nicolas Houlié[5] | Rosanne D'Arrigo[6] | James W. Kirchner[1,5] | Paolo Cherubini[1,6,7]

[1]WSL Swiss Federal Institute for Forest, Snow and Landscape Research, Birmensdorf, Switzerland

[2]Department of Geography, University of Zurich, Zürich, Switzerland

[3]Laboratory of Ion Beam Physics, ETH Zürich, Zürich, Switzerland

[4]PSI Paul Scherrer Institute, Villigen, Switzerland

[5]Department of Earth Systems Science, ETH Zürich, Zürich, Switzerland

[6]Lamont-Doherty Earth Observatory, Columbia University, Palisades, New York, USA

[7]Department of Forest and Nature Conservation, Faculty of Forestry, University of British Columbia, Vancouver, British Columbia, Canada

**Correspondence**
Paolo Cherubini, WSL Swiss Federal Institute for Forest, Snow and Landscape Research, Birmensdorf CH-8903, Switzerland.
Email: paolo.cherubini@wsl.ch

**Funding information**
Swiss National Science Foundation, Grant/Award Number: 205321_143479; University of Palermo

## Abstract

Early detection of volcanic eruptions is of major importance for protecting human life. Ground deformation and changes in seismicity, geochemistry, petrology, and gravimetry are used to assess volcanic activity before eruptions. Studies on Mt. Etna (Italy) have demonstrated that vegetation can be affected by pre-eruptive activity before the onset of eruptions. During two consecutive years before Mt. Etna's 2002/2003 flank eruption, enhanced vegetation index (NDVI) values were detected along a distinct line which later developed into an eruptive fissure. However, the mechanisms by which volcanic activity can lead to changes in pre-eruption tree growth processes are still not well understood. We analysed $\delta^{13}$C, $\delta^{18}$O and $^{14}$C in the rings of the survived trees growing near to the line where the pre-eruptive increase in NDVI was observed in order to evaluate whether the uptake of water vapour or fossil volcanic $CO_2$ could have contributed to the enhanced NDVI. We found a dramatic decrease in $\delta^{18}$O in tree rings formed before 2002/2003 in trees close to the eruption fissure, suggesting uptake of volcanic water by trees during pre-eruptive magma degassing. Moist conditions caused by outgassing of ascending magma may also have led to an observed reduction in tree-ring $\delta^{13}$C following the eruption. Furthermore, only ambiguous evidence for tree uptake of degassed $CO_2$ was found. Our results suggest that additional soil water condensed from degassed water vapour may have promoted photosynthesis, explaining local increases in NDVI before the 2002/2003 Mt. Etna flank eruption. Tree-ring oxygen stable isotopes might be used as indicators of past volcanic eruptions.

## 1 | INTRODUCTION

Early detection of precursors to volcanic eruptions can help prevent damage to infrastructure and loss of life. Ground deformation and changes in seismicity, geochemistry, petrology and gravimetry are regularly used to assess volcanic activity before eruptions (e.g., McNutt, 1996; Sicali et al., 2015). Ascending magma in volcanoes can be revealed by surface deformation, micro-seismicity and the release of volcanic gases (Aubert, 1999; Farrar et al., 1995; Sparks et al., 2012). Therefore, volcanic activity is often detected by remote sensing techniques that detect changes in geochemical and geophysical parameters related to volcanic activity, such as gas emissions and hydrological variations (e.g., Andronico et al., 2009). Studies on Mt. Etna (Italy) using satellite imagery detected pre-eruptive anomalies in the near-infrared reflectance Normalized Difference in Vegetation Index (NDVI; Huang et al., 2021), demonstrating that vegetation can also be locally affected by pre-eruptive activity months or years before the onset of eruptions (De Carolis et al., 1975; Houlié et al., 2006). An enhanced vegetation index (NDVI) was detected along a distinct line, which later developed into an eruptive fissure, over two consecutive years before the 2002/2003 flank eruption on Mt. Etna (Houlié et al., 2006). However, the processes leading to local changes in pre-eruption tree growth related to volcanic activity are still not well understood.

Analyses of climate–tree growth relationships on Mt. Etna indicate that changes in temperature and water availability are unlikely to have caused the abovementioned NDVI anomaly, suggesting that it may have arisen from other, as yet unknown, factors related to volcanic activity (Seiler, Kirchner, et al., 2017). Regarding post-eruption effects, it has been shown that photosynthetic rates of trees can be enhanced on a global scale by an increase in diffuse light caused by volcanic aerosols after eruptions (e.g., Gu et al., 2003). On the other hand, many studies have shown growth decreases after major volcanic eruptions (e.g., Büntgen et al., 2005; Krakauer & Randerson, 2003). Tree growth before, during and after eruption may be affected by pre-eruption volcanic processes and is therefore potentially useful for investigating both pre- and post-eruption effects (e.g., Seiler, Houlié, et al., 2017). After formation, growth rings remain unaltered and are stored within the plant structure, providing annually resolved archives that have been widely used to reconstruct past climatic conditions and events (e.g., Seiler, Kirchner, et al., 2017), but rarely used to study past volcanic processes and activities.

Stable carbon isotopes ($^{13}$C and $^{12}$C) in tree rings reflect a combination of atmospheric conditions and tree physiological responses to environmental changes (Francey & Farquhar, 1982) and are frequently used to study stomatal activity, water-use efficiency and the influence of climatic conditions over time (McCarroll & Loader, 2004; Shestakova & Martinez-Sancho, 2021). Stable oxygen isotope ratios ($^{18}$O and $^{16}$O) have been widely used to assess leaf transpiration, climate variability and also water sources (e.g., precipitation, soil water or groundwater) used by trees (Ehleringer & Dawson, 1992; Leonelli et al., 2017). The analysis of these isotope ratios and their variation over time can therefore provide information about some characteristics of past volcanic eruptions. Studies of the influence of volcanic eruptions on $\delta^{18}$O and $\delta^{13}$C in tree-ring cellulose found that the ejection of large amounts of dust reduced solar radiation and air temperatures and consequently led to higher atmospheric humidity, thus decreasing $\delta^{18}$O values of tree-ring cellulose (Battipaglia et al., 2007). Environmental changes that impact stomatal conductance and photosynthesis rates can also change the $\delta^{13}$C in tree-ring cellulose. For example, a change in water supply that would improve the water availability could lead to increased transpiration rates, while in contrast, any physical damage could impact the photosynthetic rates (Saurer & Cherubini, 2021). Recently, changes in $\delta^{18}$O of groundwater preceding earthquakes on Iceland were reported (Skelton et al., 2014), suggesting that seismic activity recorded before and during volcanic eruptions may induce injection of magmatic water into groundwater.

In addition to stable isotopes, radioactive isotopes are useful tracers of the photosynthetic uptake of fossil carbon dioxide, which is free of $^{14}$C and therefore can be detected by radiocarbon analysis (e.g., Donders et al., 2013; Saurer et al., 2003). Radioactive $^{14}$C decays with a half-life of 5,700 ± 30 years (Godwin, 1962). Thus, fossil $CO_2$ of geological origin (>60,000 years old) is consequently $^{14}$C-dead. Therefore, radiocarbon measurements on tree rings not only enable one to determine the $^{14}$C age of the wood but also allow the detection of fossil $CO_2$ uptake by the trees. Studies of $^{14}$C in tree rings have shown that trees growing near natural $CO_2$ springs take up fossil $CO_2$ and their growth increments contain lower levels of $^{14}$C (Donders et al., 2013; Saurer et al., 2003), and Sulerzhitzky (1971) found that trees growing near volcanoes also contain lower levels of $^{14}$C. Moreover, $^{14}$C measurements in tree rings have been used to assess uptake of fossil carbon by trees exposed to elevated concentrations of $CO_2$ from fossil fuel combustion from urban traffic (Battipaglia et al., 2010; Stuiver & Quay, 1981). Degassing of $CO_2$ from volcanic vents and its effect on trees have been demonstrated at Mammoth Mountain in California (Cook et al., 2001), where reduced $^{14}$C levels were found in growth rings of trees, demonstrating that these trees were taking up volcanic $CO_2$.

Here, we analyse isotopic ratios of $\delta^{13}$C and $\delta^{18}$O, as well as the content of $^{14}$C, in tree rings of trees growing near an eruptive fissure on Mt. Etna, along which Houlié et al. (2006) observed a pre-eruptive increase in NDVI. We aim to (1) see if there is an influence of the volcanic processes prior to the 2002/2003 eruption on isotopic ratios in tree rings formed before, during or after the eruption and (2) evaluate whether the uptake of fossil, pre-eruptive volcanic $CO_2$ (Allard et al., 1991) could have contributed to the enhanced NDVI.

## 2 | MATERIALS AND METHODS

### 2.1 | Study area

Mt. Etna is located in the northeastern part of Sicily (Italy) and, with an elevation of 3,330 m a.s.l., is the highest stratovolcano in western Europe (Branca et al., 2008). Being close to the Mediterranean Sea, it has a strongly maritime climate on its eastern slope and drier conditions on its western side. As a result of such variability in climatic and site conditions, the vegetation on Mt. Etna's slopes is very diverse (e.g., del Moral & Poli Marchese, 2010). The lower slopes feature settlements and agriculture, predominantly orchards, whereas intermediate elevations (1,000–1,600 m a.s.l.) are covered by forest stands dominated by European beech (*Fagus sylvatica* L.) and at higher elevations (1,600–1900 m a.s.l.) by Corsican black pine (*Pinus nigra* J.F. Arnold). A few undisturbed forest stands still exist at higher elevations, mostly on the northern side of the mountain. The highest vegetation belt is disturbed in its uphill development by frequent lava flows (17 since 1950; Global Volcanism Program, 2013) originating from lateral flank eruptions (Certini et al., 2001; Egli et al., 2012), or wildfires caused by other volcanic processes, such as degassing and heat radiation through small vents (Allard, 1997; Aubert, 1999). Predominant soil types are Regosols, Andosols and Cambisols, with different characteristics depending on the age of the lava flows and volcanic deposits on which they developed (Egli et al., 2008; Lulli, 2007). Natural and anthropogenic processes, i.e., wildfires, lava flows, avalanches and logging, have heavily impacted the forests on the flanks of Mt. Etna over past millennia.

## 2.2 | Sampling

Given the fact that the trees growing along the line showing enhanced NDVI values did not survive the eruption, two cores were collected at 1.30 m tree height from each of 51 *P. nigra* trees growing in the immediate vicinity to the 2002/2003 eruptive fissure near Piano Provenzana, at 1,700–1,800 m a.s.l. on the north-eastern flank of Mt. Etna (Figure 1). A 0.5 cm-diameter corer with a three-threaded auger (Haglof Inc., Sweden) was used to core the trees.

## 2.3 | Ring-width measurements

We measured ring width to the nearest 0.01 mm using a Leica Wild M32 binocular (Leica, Germany) with 25–50X magnification, coupled to a LINTAB measuring device and a computer running the TSAP (Time Series Analysis Program) software (RinnTech, Heidelberg, Germany). Ring-width measurements were crossdated visually and statistically using TSAP and Cofecha (50-year segments with a 25-year overlap) to ensure that all rings were correctly dated and no rings were missing (e.g., Seiler, Kirchner, et al., 2017).

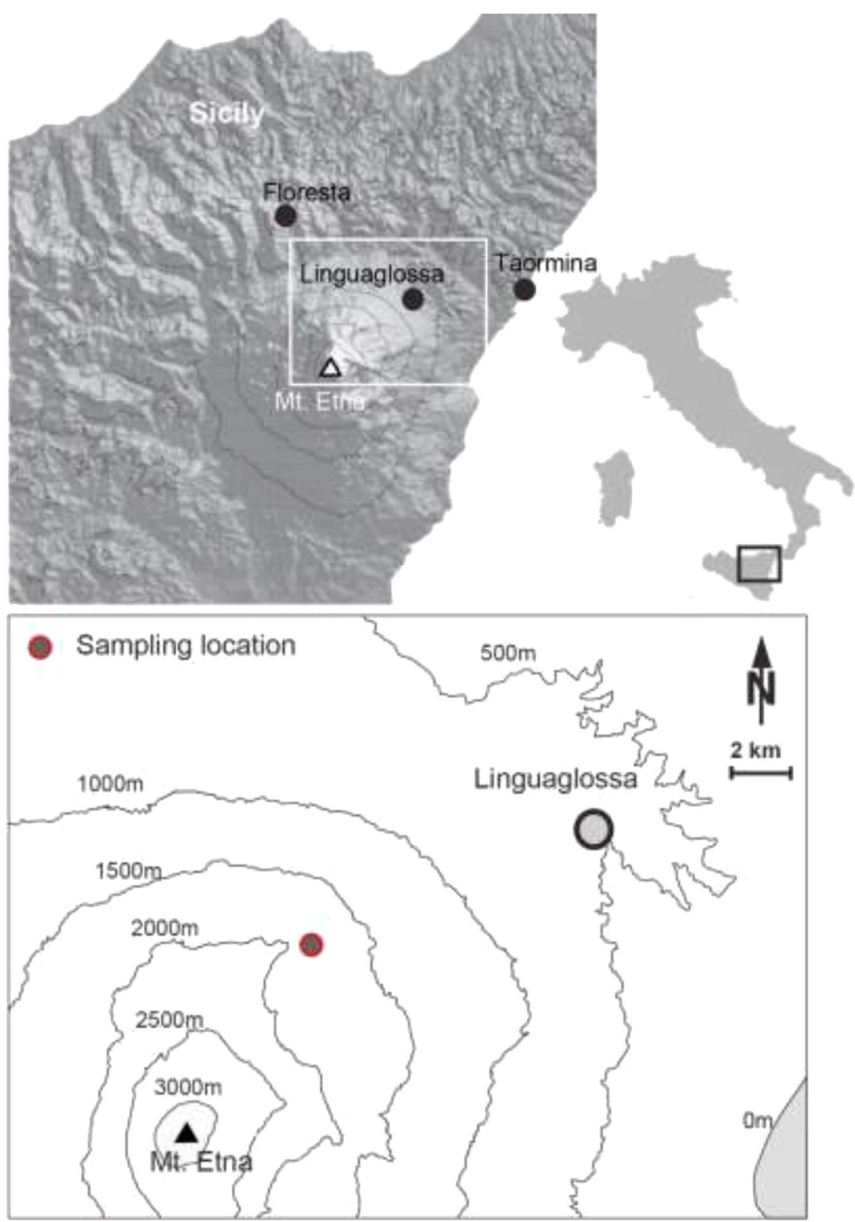

**FIGURE 1** Sampling location (red circle) along the 2002/2003 eruptive fissure on the northeastern flank of Mt. Etna at an elevation range of 1,600 to 1,850 m a.s.l. The map of Italy was created using the program R (Version 3.1.3; URL: http://www.R-project.org/), the topographic map showing Sicily was created using Generic Mapping Tools (Version 5.2.1; URL: http://gmt.soest.hawaii.edu/) and the lower panel was taken from Egli et al. (2008) and modified

## 2.4 | Sample preparation

Five tree cores from five different trees were taken to measure stable isotope compositions in the tree rings. Wood samples of tree rings between 1977 and 2006 were prepared for stable isotope analyses.

From four other tree cores, in order to have enough material, wood samples of five consecutive annual rings (1999–2003) overlapping the time of eruption, and an additional five consecutive years (1992–1996, i.e., used as control rings formed when no eruption occurred at that location), were selected for radiocarbon measurement. Individual annual rings of selected cores were separated and cut into pieces under a binocular with a 25X magnification using a standard laboratory scalpel. After each cutting sequence, the material of the separated ring was stored in an Eppendorf tube, and the workspace and instruments were cleaned using a vacuum cleaner and wiper tissues impregnated with ethanol/distilled water to prevent mixture of wood fragments from different ring samples.

## 2.5 | Stable isotope analyses

Cellulose extraction followed the procedure of Boettger et al. (2007), in which samples from five tree cores (corresponding rings) were homogenized in an ultra-centrifugal mill at 16,000 rotations per minute. Ten milligrammes of the resulting wood powder from each tree-ring sample were enclosed in Teflon filter bags and washed twice with a 5% sodium hydroxide (NaOH) solution for 2 h at 60°C. These washing runs remove oils, tannins, fats, hemi-cellulose and resins. To eliminate any remaining lignin, we immersed the samples in a 7% solution of sodium chlorite ($NaClO_2$) at 60°C for a total of 30 h. Samples were then dried in the oven at 60°C overnight.

Following Woodley et al. (2012) and Weigt et al. (2015), the carbon and oxygen stable isotopic composition of the cellulose was measured on CO gas produced by pyrolysis in a TC/EA at 1,420°C, coupled with an isotope-ratio mass spectrometer (delta Plus XP, Thermo, Bremen, Germany), on 1.0–1.2 mg of wood cellulose wrapped in silver cups, in a continuous flow of pure He gas. The isotopic ratios ($R = {}^{13}C/{}^{12}C$ or ${}^{18}O/{}^{16}O$) of each sample were measured against a reference CO gas and expressed in the $\delta$-notation, where

$$\delta(‰) = \{(R_{sample}/R_{standard}) - 1\} \times 1,000.$$

The mass spectrometer was calibrated to international V-PDB and V-SMOW standards for C and O isotopes, respectively, as well as internal laboratory standards and provided a precision of 0.2‰ for $\delta^{13}C$ and 0.3‰ for $\delta^{18}O$.

Due to the release of isotopically light carbon from burning of fossil fuel over the industrial period, $\delta^{13}C$ of atmospheric $CO_2$ has decreased over time. Our $\delta^{13}C$ data were therefore corrected for this so-called '$^{13}C$ Suess-effect' (Keeling, 1979) by adding values provided by McCarroll and Loader (2004), as recommended in the literature (Belmecheri & Lavergne, 2020).

## 2.6 | Radiocarbon—Preparation methods

After Soxhlet treatment (described below), the wood samples were prepared for radiocarbon analysis using two different methods, i.e., the cellulose-extraction method (Němec et al., 2010), and the acid–base–acid (ABA) washing method (Hajdas, 2008), to compare differences in $^{14}C$ content between the two methods. ABA-prepared samples were measured to verify that the values obtained analysing the samples prepared by the cellulose-extraction method were not due to contamination (Hajdas et al., 2017). In order to collect sufficient amounts of material for both methods and obtain robust $^{14}C$ measurements, we pooled samples on an annual basis, combining sample material of the same growth year of the cores taken from the four trees selected for radiocarbon measurements.

### 2.6.1 | Soxhlet treatment

Earlier studies addressed the importance of standard Soxhlet treatment prior to radiocarbon dating to remove $^{14}C$ from organic matter containing possible contamination with waxes and resins (Hajdas, 2008). After being placed in a Soxhlet apparatus, samples were immersed sequentially in circulating condensates of hexane, acetone and ethanol for one hour each, following the technique described by Hajdas (2008).

### 2.6.2 | Cellulose extraction

We extracted cellulose from our samples using a five-step process, during which each sample was immersed in 5 ml of 4% sodium hydroxide (NaOH) and hydrogen chloride (HCl) at 75°C alternately for 10, 1, 1.5 and 1 h, respectively. After each step, the samples were rinsed with distilled water until they were pH-neutral (measured with pH test strips). During the fifth and final step, samples were immersed in 5 ml of 5% sodium chlorite ($NaClO_2$) mixed with five drops of 4% HCl at 75°C for 2 h prior to being put into an ultrasonic bath for 15 min. Finally, samples were washed neutral in distilled $H_2O$ at 60°C and oven-dried (Němec et al., 2010).

### 2.6.3 | ABA method

Designed to remove contamination of samples by humic acids and carbonates (De Vries and Barendsen, 1954), the ABA method is the standard chemical pre-treatment of organic matter for radiocarbon dating (Hajdas, 2008; Hajdas et al., 2017; Southon & Magana, 2010). Soxhlet-treated samples were first washed in an acid solution to remove carbonate contamination from the sample surface. The samples were then washed with distilled water before being immersed into 0.1 M sodium hydroxide (NaOH) for 1 h to dissolve humic acids. After rinsing to neutral pH, samples were then immersed in an acid bath to remove any carbonates that may have precipitated from modern atmospheric $CO_2$ dissolved in the NaOH solution. Finally, all samples were washed to pH neutral and dried.

## 2.7 | Graphitization and $^{14}C$ measurements

Prior to the AMS analysis, all the samples were combusted, and the pure $CO_2$ was reduced to graphite. An equivalent of 1 mg of carbon (approximately 3.5 mg of cellulose) was weighed into the tin boats for graphitization using the AGE system (Wacker et al., 2010). AMS analysis was performed on the miniaturized radiocarbon dating system (MICADAS; Synal et al., 2007) at the laboratory of Ion Beam Physics of ETH Zürich.

## 2.8 | Meteorological data

We calculated the relationship between climate variability and isotopic ratios by using an interpolated monthly temperature and precipitation dataset for the Mt. Etna region provided by the CRU (Climatic Research Unit), University of East Anglia, Norwich, U.K. (e.g., Mitchell et al., 2004).

## 2.9 | Mixed effect models

Using a forward–backward approach based on the Akaike Information Criterion, we calculated mixed effect models, incorporating meteorological variables, which explained the highest variance of stable isotope ratios (Cook & Kariukstis, 1990). To avoid overfitting, we tested the explanatory variables for collinearity using variance inflation factor analysis (Kutner et al., 2005). The models were calibrated using the timespan from 1977 to 2006.

## 3 | RESULTS

### 3.1 | Tree-ring chronologies

The mean raw ring-width chronology is displayed in Figure 2. The series intercorrelation of 0.5, as an indicator of the common growth signal, demonstrates that these tree-ring series match and crossdate well that no tree rings are missing and that the calendar ages (AD) of the tree rings are correctly attributed. Tree-ring width chronology statistics are shown in Table 1.

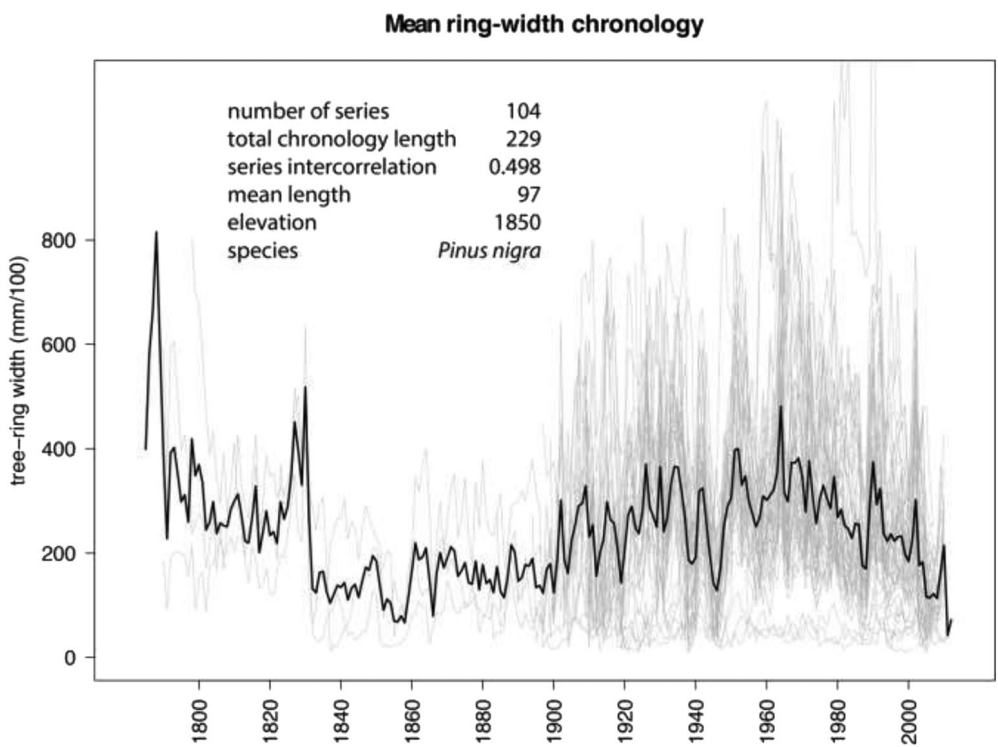

**FIGURE 2** Tree-ring width of single trees (grey) and mean ring-width chronology of the site (black)

**TABLE 1** Descriptive statistics of the mean ring-width chronology displaying number of trees, total length (years) of chronology, series intercorrelation, mean sample length (years), elevation (m a.s.l.) and species

| Site | No. of trees | Total length | ser.interc. | Mean length | Elevation | Species |
| --- | --- | --- | --- | --- | --- | --- |
| Piano Provenzana | 51 | 228 | 0.498 | 91.7 | 1,850 | *Pinus nigra* |

Abbreviations: ser.interc., measure of common growth signal in the chronology.

**TABLE 2** Stable isotope ratios (given in ‰) of $\delta^{18}O$ (30.01 to 34.94) and $\delta^{13}C$ (−24.70 to −20.33) of single trees (T1–T5) and average ratios

| Year | T1_$\delta^{18}$O | T1_$\delta^{13}$C | T2_$\delta^{18}$O | T2_$\delta^{13}$C | T3_$\delta^{18}$O | T3_$\delta^{13}$C | T4_$\delta^{18}$O | T4_$\delta^{13}$C | T5_$\delta^{18}$O | T5_$\delta^{13}$C | avg_$\delta^{18}$O | avg_$\delta^{13}$C |
|---|---|---|---|---|---|---|---|---|---|---|---|---|
| 1977 | 33.93 | −23.23 | | | 32.85 | −22.66 | 33.05 | −23.51 | 33.51 | −24.06 | 33.33 | −23.37 |
| 1978 | 33.62 | −22.14 | | | 33.70 | −22.04 | 32.76 | −21.83 | 32.87 | −21.70 | 33.24 | −21.93 |
| 1979 | 34.10 | −23.17 | | | 33.75 | −22.41 | 33.61 | −23.29 | 33.57 | −23.09 | 33.76 | −22.99 |
| 1980 | 33.65 | −23.38 | | | 32.74 | −23.52 | 33.29 | −23.80 | 33.18 | −22.58 | 33.22 | −23.32 |
| 1981 | 34.53 | −22.56 | | | 32.74 | −22.22 | 34.59 | −24.39 | 33.78 | −21.12 | 33.91 | −22.58 |
| 1982 | 34.76 | −22.48 | | | 33.58 | −22.32 | 34.94 | −22.95 | 34.10 | −21.11 | 34.35 | −22.22 |
| 1983 | 33.79 | −23.28 | 33.15 | −22.19 | 33.76 | −22.11 | 33.82 | −23.72 | 33.48 | −23.20 | 33.60 | −22.90 |
| 1984 | 33.15 | −22.79 | 32.84 | −20.63 | 34.28 | −22.65 | 33.20 | −21.80 | 32.38 | −22.24 | 33.17 | −22.02 |
| 1985 | 33.70 | −22.37 | 33.67 | −22.25 | 33.85 | −21.59 | 34.36 | −21.18 | 33.53 | −22.23 | 33.82 | −21.92 |
| 1986 | 33.43 | −22.72 | 32.68 | −22.31 | 34.48 | −21.89 | 32.96 | −22.25 | 32.45 | −22.22 | 33.20 | −22.28 |
| 1987 | 32.55 | −22.14 | 32.23 | −22.33 | 33.30 | −21.88 | 32.35 | −22.15 | 31.83 | −22.85 | 32.45 | −22.27 |
| 1988 | 32.46 | −22.92 | 31.99 | −22.66 | 32.55 | −22.89 | 33.41 | −21.88 | 32.66 | −21.05 | 32.61 | −22.28 |
| 1989 | 32.68 | −22.97 | 33.23 | −23.18 | 32.77 | −22.31 | 32.87 | −23.77 | 32.43 | −21.51 | 32.80 | −22.75 |
| 1990 | 32.98 | −21.98 | 32.93 | −22.10 | 32.86 | −22.31 | 33.64 | −21.41 | 33.08 | −21.77 | 33.10 | −21.91 |
| 1991 | 33.08 | −22.65 | 32.57 | −22.22 | 33.77 | −21.90 | 33.43 | −22.59 | 32.62 | −21.42 | 33.09 | −22.16 |
| 1992 | 33.30 | −22.69 | 32.71 | −22.29 | 32.99 | −22.50 | 32.74 | −22.05 | 32.22 | −22.36 | 32.79 | −22.38 |
| 1993 | 32.99 | −21.32 | 32.34 | −21.96 | 32.66 | −22.44 | 33.53 | −21.64 | 32.94 | −23.34 | 32.89 | −22.14 |
| 1994 | 32.21 | −22.96 | 32.43 | −21.69 | 32.63 | −20.33 | 33.91 | −24.07 | 32.69 | −22.67 | 32.77 | −22.34 |
| 1995 | 31.83 | −23.26 | 31.95 | −23.35 | 34.04 | −21.88 | 33.29 | −23.51 | 32.56 | −23.55 | 32.73 | −23.11 |
| 1996 | 32.05 | −23.44 | 32.81 | −23.52 | 33.04 | −22.52 | 33.22 | −24.43 | 32.18 | −23.02 | 32.66 | −23.39 |
| 1997 | 33.16 | −22.66 | 33.14 | −22.85 | 32.11 | −22.70 | 33.29 | −22.65 | 31.70 | −23.97 | 32.68 | −22.97 |
| 1998 | 32.50 | −22.97 | 33.95 | −22.56 | 33.65 | −21.48 | 34.03 | −23.15 | 32.08 | −21.13 | 33.24 | −22.26 |
| 1999 | 32.66 | −23.56 | 32.77 | −22.14 | 32.39 | −22.18 | 33.18 | −22.22 | 31.33 | −22.52 | 32.47 | −22.52 |
| 2000 | 31.43 | −22.86 | 32.68 | −22.50 | 32.86 | −22.48 | 33.08 | −22.62 | 31.73 | −22.09 | 32.35 | −22.51 |
| 2001 | 33.18 | −22.40 | 33.93 | −21.68 | 33.80 | −22.81 | 33.83 | −22.21 | 33.25 | −23.35 | 33.60 | −22.49 |
| 2002 | 32.41 | −23.63 | 33.21 | −22.90 | 32.52 | −23.07 | 33.66 | −23.49 | 32.43 | −23.55 | 32.85 | −23.33 |
| 2003 | 30.24 | −23.11 | 31.48 | −22.58 | 30.75 | −23.82 | 30.88 | −23.58 | 30.01 | −24.70 | 30.67 | −23.56 |
| 2004 | 31.77 | −23.69 | 34.08 | −21.55 | 33.56 | −23.90 | 32.93 | −23.45 | 31.06 | −23.41 | 32.68 | −23.20 |
| 2005 | 30.54 | −24.06 | 32.06 | −23.06 | 33.12 | −22.90 | 32.02 | −23.19 | | | 31.93 | −23.30 |
| 2006 | | | 33.60 | −23.151 | 32.71 | −23.47 | 33.12 | −23.79 | 32.36 | −23.78 | 32.95 | −23.55 |

*Note*: For years with missing values, there was not enough cellulose material to measure isotopic ratios.

## 3.2 | Stable isotopic ratios—$\delta^{13}C$ and $\delta^{18}O$

Tree-ring $\delta^{13}C$ measurements range from −24.70‰ to −20.33‰ and $\delta^{18}O$ from 30.01‰ to 34.94‰ (with series inter-correlations of 0.44 and 0.39, respectively; Table 2). Isotope ratios of single tree cores and stand averages are given in Table 2. We performed a running-correlation analysis using an 11-year window to analyse changes in synchronicity between $\delta^{13}C$ and $\delta^{18}O$ ratios over time. The results (Figure 3) show that carbon and oxygen isotope ratio variations are poorly correlated between 1982 and 1986 ($|\rho| < 0.05$; $P > 0.9$) but much more strongly correlated between 1989 and 2006 ($0.31 < \rho < 0.78$; $P < 0.05$).

We next applied a linear detrending technique to our stable isotope data to remove any long-term positive or negative trends and to distinguish between correlations influenced by high- and low-frequency variability in the data (Cook & Kariukstis, 1990).

Based on Spearman rank correlations (Figure 4) for the pre-eruption period 1977–2002, the raw and detrended tree ring $\delta^{13}C$ values are positively correlated with temperatures in September ($\rho = 0.441$ and 0.412; $P < 0.05$) and October ($\rho = 0.403$ and 0.418; $P < 0.05$). The raw $\delta^{13}C$ measurements are negatively correlated with precipitation in June ($\rho = -0.393$; $P < 0.05$), and the detrended $\delta^{13}C$ measurements are negatively correlated with precipitation in August ($\rho = -0.434$; $P < 0.05$). Both raw and detrended $\delta^{13}C$ measurements revealed a common response to climate parameters, indicating that the relationship is caused by short-term variability.

The raw tree-ring $\delta^{18}O$ is negatively correlated with temperature in August ($\rho = -0.41$; $P < 0.05$), in contrast to the detrended tree-ring $\delta^{18}O$, which is positively correlated with temperature in June ($\rho = 0.565$; $P < 0.05$) and negatively correlated with precipitation in June ($\rho = -0.42$; $P < 0.05$).

Thus, the contrasting correlations between raw and detrended $\delta^{18}O$ stable isotope ratios with climate are clearly caused by a long-term trend. The long-term decreasing trend in the oxygen stable isotope ratios is inversely correlated with a generally increasing trend in summer temperatures, leading to the negative correlation with August temperature. Based on those variables significantly correlating with short-term variations of $\delta^{18}O$, i.e., temperature and precipitation in June, the strongly reduced $\delta^{18}O$ in 2003 cannot be explained by either of these two variables, which do not significantly deviate from their long-term average (t test; $P > 0.05$) for that particular year.

Further, the mixed-effect model explains 50% (adjusted $R^2$; $P < 0.001$) of $\delta^{13}C$ variability, demonstrating that the ratio of heavy to light stable carbon isotopes is predominantly influenced by climate, also before and during the eruption of 2002/2003 (Figure 5).

In contrast, the mixed-effect model explains 24% (adjusted $R^2$; $P < 0.05$) of $\delta^{18}O$ variability between 1977 and 2006. Moreover, the strong reduction in 2003 exceeds the two-standard-deviation limit and is not explained by the model, suggesting that climate is unlikely to have caused this reduction (Figure 6). This negative

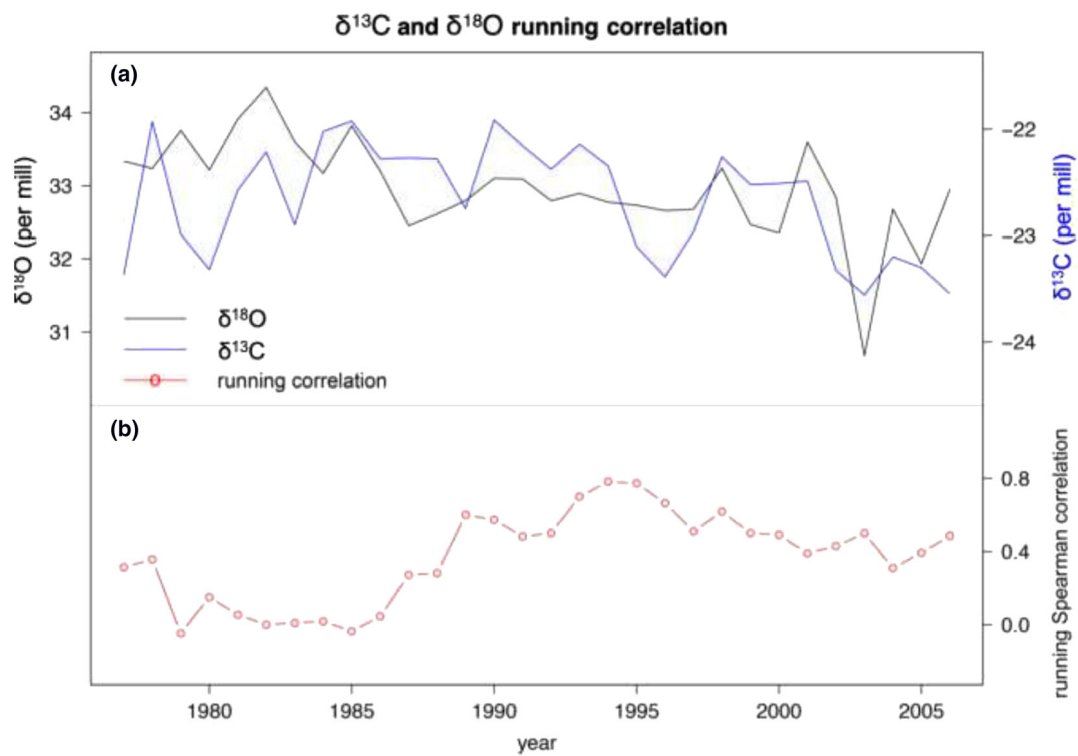

**FIGURE 3** Average $\delta^{13}C$ and $\delta^{18}O$ chronologies (a) and Spearman rank running correlation $\rho$ values from 1977 to 2006 between the two isotopes calculated with a window of 11 years (b)

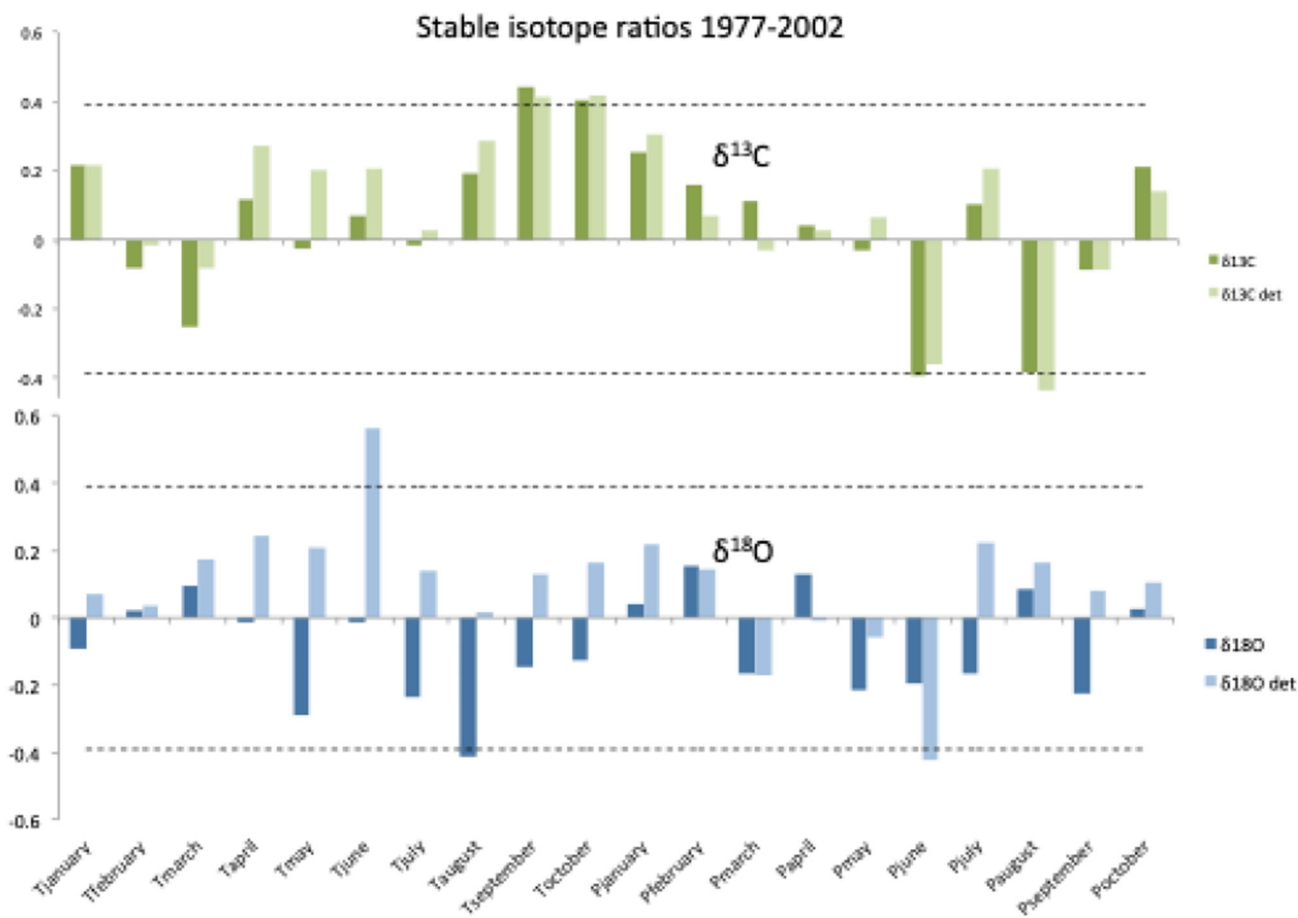

**FIGURE 4** Spearman rank correlations between stable isotope ratios and climate variables, where T = temperature and P = precipitation. The 95% significance levels are indicated by dashed lines. "det" stands for detrended

anomaly in $\delta^{18}O$ is most pronounced when considering detrended data (Figure 6b,d).

## 3.3 | Radiocarbon—$^{14}C$

Radiocarbon dating of wood material is reported as fraction modern ($F^{14}C$). Results are summarized in Table 3. The $^{14}C$ measurements in samples prepared by cellulose extraction indicate that for all years in the control period (1992–1996), as well as in the eruptive period (1999–2003), there was no evidence that trees had taken up fossil $CO_2$. The $^{14}C$ measurements in samples prepared with ABA also showed no depleted $^{14}C$ signal during the control period, but we found lower $^{14}C$ ratios in tree rings for 2001 and 2002, immediately before the eruptive period.

We measured a second set of ABA-prepared samples to obtain additional measurement points and to verify that the reduced values before the eruption were not due to contamination. The two runs are consequently referred to as ABA1 and ABA2 (Table 3). One of the ABA2 samples exhibits a slight (roughly 1%) depletion of $^{14}C$ in 2001, but the other four ABA2 samples (from 1999, 2000, 2002 and 2003) show no $^{14}C$ depletion (Figure 7).

## 4 | DISCUSSION

### 4.1 | Stable isotopic ratios: $\delta^{13}C$ and $\delta^{18}O$

The generally positive relationship between carbon isotope ratios and temperature and the negative relationship with precipitation are in agreement with theoretical understanding and previously reported influences of climate on $\delta^{13}C$, reflecting the positive influence of moisture on photosynthesis and stomatal opening (Cherubini et al., 2021). Under drought, i.e., low precipitation, plants need to limit water loss by closing their stomata, which results in reduced intercellular $CO_2$ concentrations, reduced discrimination against $^{13}C$ and subsequently higher $\delta^{13}C$ (Francey & Farquhar, 1982). There were only minor differences between the raw and the detrended measurements, demonstrating that there was little influence of low-frequency variability on the isotope-climate correlations and hence no substantial long-term trend in the data. During and after the eruption in 2002/2003, the $\delta^{13}C$ values were somewhat lower than in the years before and could be explained mainly by climate variability. Taking into account that the $\delta^{13}C$ values of deep-origin $CO_2$ in the region are much higher, i.e., less negative (Camarda et al., 2020), we may exclude a possible influence of

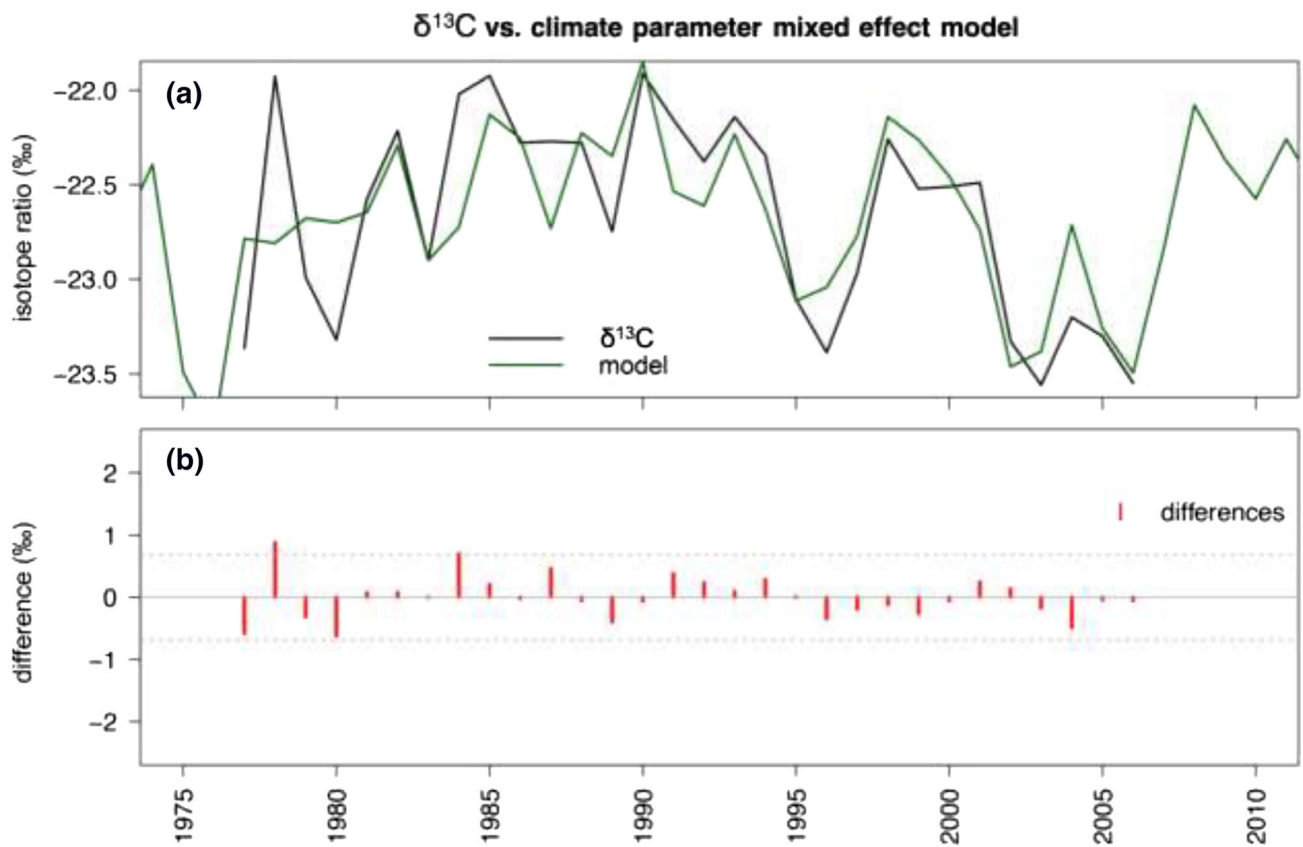

**FIGURE 5** Raw $\delta^{13}C$ variability and the mixed-effect models are shown in the top panel (a). The deviation between measurements and models (measurements–model) is shown in the bottom panel (b)

the ongoing volcanic activity on the carbon isotope ratio within the trees.

In contrast, the oxygen isotope correlations revealed a divergence between raw and detrended data. While there is a negative relationship between raw $\delta^{18}O$ and temperature (the opposite of commonly observed correlations; e.g., Reynolds-Henne et al., 2007), the detrended $\delta^{18}O$ data are positively correlated with temperature and negatively correlated with precipitation, thus in better agreement with expectations and previous studies (e.g., Cherubini et al., 2021). The contrasting correlations with climate variables between raw and detrended stable oxygen isotope values clearly stem from low-frequency variability, in which the long-term negative trend in $\delta^{18}O$ (Figure 3) is inversely correlated with an increasing temperature trend. The negative isotope trend could be related, for example, to slowly increasing rooting depth with tree age and a concomitant shift in source water (Esper et al., 2010). However, after this long-term trend is removed, the detrended $\delta^{18}O$ data display the commonly observed correlations with temperature, which much more plausibly reflect the relationship between high-frequency variations in $\delta^{18}O$ and temperature, representing the direct influence of climate on tree-ring $\delta^{18}O$.

Positive $\delta^{18}O$–temperature and negative $\delta^{18}O$–precipitation correlations in the detrended $\delta^{18}O$ data may be explained by reduced stomatal conductance with higher temperatures and lower humidity resulting in strong isotopic enrichment of leaf water (Saurer & Cherubini, 2021). This commonly observed relationship is further strengthened by the isotopic signature of the water source (usually precipitation), which also depends on temperature due to fractionation in the hydrological cycle (Coplen et al., 2000).

The most striking finding is that the $\delta^{18}O$ values in 2003 are strongly negative compared to all other years preceding the event (Figure 3). Positive correlations between $\delta^{18}O$ and temperature in the pre-eruption period suggest that either very low temperatures in 2003 could have caused this reduced heavy oxygen signal or that the water source used by the trees during that time was depleted in $^{18}O$ at our sample location on Mt. Etna. However, temperatures in June, which show the strongest correlation with $\delta^{18}O$, were even above average in 2003, showing that the ratio was not reduced by low temperatures. Thus, the very low $\delta^{18}O$ values most likely were induced by a change in the source water, possibly linked to seismic activity. Such an effect, unrelated to climate, could have been caused by deep ground water rising up to the surface. Long-established confined groundwaters migrate to the surface when new hydraulic pathways open during seismic deformations, as shown by depleted $\delta^{18}O$ values in water found in India (Reddy & Nagabhushanam, 2012), Japan (Hosono et al., 2018, 2020; Onda et al., 2018) and in Iceland (Skelton et al., 2019) a few months before and after earthquakes. Degassing of volcanic water vapour could have provided an additional water source, causing a localized increase in humidity close to the volcanic vent (Allen et al., 2002), thereby avoiding stomatal closure and altering isotopic ratios of the water source leading to a reduction in δ$^{18}$O. The decrease in δ$^{18}$O values may have been induced by factors unrelated to climate. In 2003, the trees may either (1) have taken up an increased proportion of groundwater that was low in δ$^{18}$O compared to precipitation and surface waters (Balsch & Bryson, 2007; Coplen et al., 2000; Giggenbach & Soto, 1992; Kalbus et al., 2006; Onda et al., 2018) or (2) have incorporated soil moisture that condensed from volcanic vapour/gases (negative in δ$^{18}$O; Ohwada et al., 2003; D'Alessandro et al., 2004) which provided an unusual water source. The negative δ$^{18}$O of volcanic waters may be related to oxygen exchange between water and $CO_2$ gas (Caliro et al., 2005).

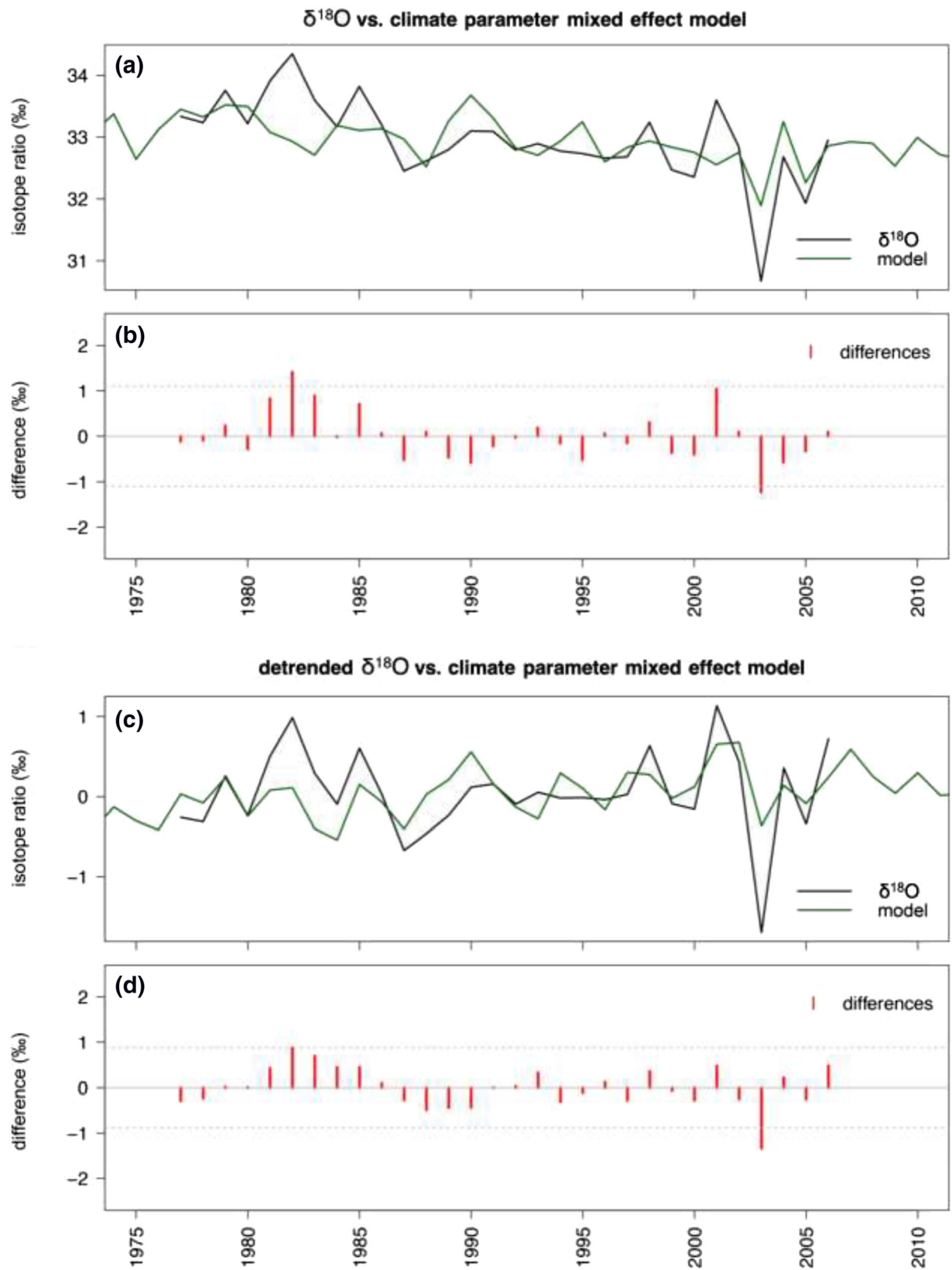

**FIGURE 6** Raw and detrended $\delta^{18}O$ variability and the mixed effect models are shown in the top panels (a and c). The deviations between measurements and models (measurements−model) are shown in the bottom panels (b and d)

**TABLE 3** Radiocarbon measurements of single trees, expressed as the fraction of modern atmospheric $^{14}$C (F$^{14}$C)

| Dendro Age | C14 age BP | ±1σ | F$^{14}$C | ±1σ | δC13 ‰ | ±1σ | mg C | C/N | %C | Preparation Method 1 | Method 2 |
|---|---|---|---|---|---|---|---|---|---|---|---|
| 2003 | −600 | 24 | 1.078 | 0.00 | −24.8 | 1.1 | 0.98 | 59.26 | | Soxhlet | Cellulose |
| 2003 | −624 | 24 | 1.081 | 0.00 | −24.1 | 1.1 | 0.99 | 56.59 | | Soxhlet | Cellulose |
| 2003 | −596 | 22 | 1.077 | 0.0029 | −22.7 | 1 | 1 | 137.04 | | Soxhlet | ABA 2 |
| 2002 | −607 | 24 | 1.078 | 0.00 | −21.5 | 1.1 | 1.00 | 60.88 | | Soxhlet | Cellulose |
| 2002 | −626 | 24 | 1.081 | 0.00 | −23.0 | 1.1 | 1.00 | 53.22 | | Soxhlet | Cellulose |
| 2002 | −353 | 24 | 1.045 | 0.00 | −24.6 | 1.1 | 0.99 | 75.16 | | Soxhlet | ABA 1 |
| 2002 | −342 | 24 | 1.043 | 0.00 | −25.1 | 1.1 | 0.99 | 55.59 | | Soxhlet | ABA 1 |
| 2002 | −641 | 22 | 1.0831 | 0.003 | −23.6 | 1 | 0.99 | 125.81 | | Soxhlet | ABA 2 |
| 2001 | −688 | 24 | 1.089 | 0.00 | −23.4 | 1.1 | 0.99 | 53.32 | | Soxhlet | Cellulose |
| 2001 | −675 | 24 | 1.088 | 0.00 | −22.3 | 1.1 | 0.99 | 54.35 | | Soxhlet | Cellulose |
| 2001 | −440 | 24 | 1.056 | 0.00 | −21.9 | 1.1 | 0.99 | 63.98 | | Soxhlet | ABA 1 |
| 2001 | −470 | 25 | 1.060 | 0.00 | −25.6 | 1.1 | 1.00 | 65.02 | | Soxhlet | ABA 1 |
| 2001 | −600 | 22 | 1.0775 | 0.003 | −24.6 | 1 | 0.97 | 121.05 | | Soxhlet | ABA 2 |
| 2000 | −749 | 24 | 1.098 | 0.00 | −27.5 | 1.1 | 1.00 | 49.38 | | Soxhlet | Cellulose |
| 2000 | −727 | 24 | 1.095 | 0.00 | −22.8 | 1.1 | 0.99 | 54.07 | | Soxhlet | Cellulose |
| 2000 | −730 | 22 | 1.0951 | 0.003 | −24.3 | 1 | 0.94 | 114.64 | | Soxhlet | ABA 2 |
| 1999 | −751 | 24 | 1.098 | 0.00 | −21.1 | 1.1 | 0.98 | 59.96 | | Soxhlet | Cellulose |
| 1999 | | | | | | | | | | Soxhlet | ABA 1 |
| 1999 | −749 | 25 | 1.098 | 0.00 | −18.5 | 1.1 | 0.99 | 53.43 | | Soxhlet | Cellulose |
| 1999 | −761 | 22 | 1.0993 | 0.003 | −25.1 | 1 | 1 | 125.48 | | Soxhlet | ABA 2 |
| 1996 | −897 | 24 | 1.118 | 0.00 | −26.5 | 1.1 | 1.00 | 278.7 | 50.41 | Soxhlet | Cellulose |
| 1996 | | | 0.000 | 0.00 | 0.0 | 1.1 | 0.00 | | 0.00 | Soxhlet | Cellulose |
| 1996 | −884 | 24 | 1.116 | 0.00 | −25.9 | 1.1 | 0.99 | 234.0 | 60.68 | Soxhlet | ABA 1 |
| 1995 | −933 | 24 | 1.123 | 0.00 | −25.6 | 1.1 | 1.00 | 236.6 | 52.77 | Soxhlet | Cellulose |
| 1995 | | | 0.000 | 0.00 | 0.0 | 1.1 | 0.00 | | 0.00 | Soxhlet | Cellulose |
| 1995 | −909 | 24 | 1.120 | 0.00 | −22.1 | 1.1 | 1.00 | 212.2 | 57.03 | Soxhlet | ABA 1 |
| 1994 | −967 | 24 | 1.128 | 0.00 | −22.6 | 1.1 | 0.99 | 253.6 | 56.99 | Soxhlet | Cellulose |
| 1994 | | | 0.000 | 0.00 | 0.0 | 1.1 | 0.00 | | 0.00 | Soxhlet | Cellulose |
| 1994 | −915 | 24 | 1.121 | 0.00 | −19.2 | 1.1 | 1.00 | 212.5 | 57.88 | Soxhlet | ABA 1 |
| 1993 | −997 | 24 | 1.132 | 0.00 | −20.4 | 1.1 | 1.00 | 289.3 | 64.20 | Soxhlet | Cellulose |
| 1993 | | | 0.000 | 0.00 | 0.0 | 1.1 | 0.00 | | 0.00 | Soxhlet | Cellulose |
| 1993 | −1,002 | 24 | 1.133 | 0.00 | −20.2 | 1.1 | 0.99 | 193.6 | 59.05 | Soxhlet | ABA 1 |
| 1992 | −1,034 | 24 | 1.137 | 0.00 | −18.8 | 1.1 | 1.00 | 289.6 | 64.56 | Soxhlet | Cellulose |
| 1992 | | | 0.000 | 0.00 | 0.0 | 1.1 | 0.00 | | 0.00 | Soxhlet | Cellulose |
| 1992 | −997 | 24 | 1.132 | 0.00 | −20.5 | 1.1 | 0.98 | 202.7 | 58.07 | Soxhlet | ABA 1 |

Significant changes in oxygen isotopic ratios induced by changes in tectonic settings were also found to be related to earthquakes on Iceland (Skelton et al., 2014), suggesting that seismic activity before or during volcanic eruptions may similarly have altered groundwater composition on Mt. Etna by infiltration of volcanic water vapour

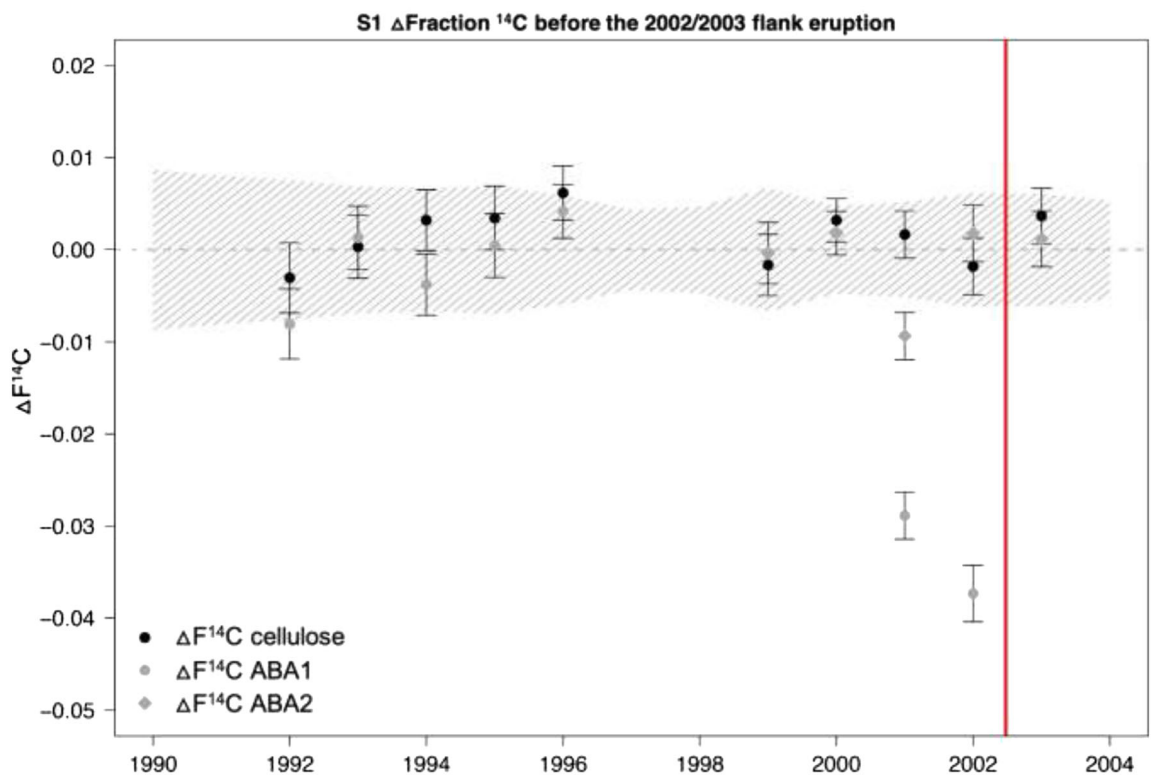

**FIGURE 7** Deviations of $\Delta F^{14}C$ values from atmospheric composition (grey dashed line), for two different preparation methods, cellulose (black dots) and ABA (grey dots and grey diamonds)

(depleted in $\delta^{18}O$; D'Alessandro et al., 2004) into shallow ground. This is supported by a study that showed volcanic activity was likely to have led to an increase in soil moisture prior to a flank eruption on Mt. Etna (Seiler, Kirchner, et al., 2017). Therefore, we argue that additional water availability provided by degassing of volcanic water vapour (depleted in $\delta^{18}O$) may have led to moist conditions during the eruption of 2002/2003, influencing stable isotope ratios in the 2003 tree rings. Under such conditions, we would expect a higher stomatal conductance and subsequently reduced $\delta^{13}C$ ratios, consistent with the measurements reported here.

## 4.2 | Radiocarbon—$^{14}C$

Applying the cellulose-extraction preparation method, we did not find any evidence that the trees incorporated $^{14}C$-depleted $CO_2$ during the eruption event of 2002/2003. However, applying the ABA preparation method, we found some negative $\Delta F^{14}C$ values, suggesting depletion of $^{14}C$ due to uptake of fossil $CO_2$, during the two consecutive years (i.e., 2001 and 2002) just before the 2002/2003 eruption. Since only three out of six ABA measurements prior to the eruption exhibited negative $\Delta F^{14}C$, there is only ambiguous evidence for tree uptake of $^{14}C$-depleted $CO_2$ during two consecutive years preceding the surface eruption.

The refilling of shallow conduits with magma can lead to pre-eruption $CO_2$ degassing and infiltration of $CO_2$ into soils (Aiuppa et al., 2013). Therefore, we hypothesize that the rift zone trees sampled here might have been exposed to $CO_2$ degassing, leading to depleted values of $^{14}C$ in the 2001 and 2002 rings prior to the eruption of 2002/2003, the exact time window during which the increased NDVI was observed along the rift (Houlié et al., 2006).

## 5 | CONCLUSIONS

Given the dramatic decrease in tree-ring $\delta^{18}O$ in immediate proximity to the eruption fissure, we conclude that it is likely that the 2002/2003 volcanic eruption impacted trees during and after the effusive eruption phase. A plausible explanation for the reduced $\delta^{18}O$ in the trees is the uptake of volcanic water during pre-eruption degassing.

Our findings suggest that volcanic water vapour reduced $\delta^{18}O$ values in tree rings after the eruption and that moist conditions caused by outgassing of ascending magma may have led to a reduction in $\delta^{13}C$ values in tree rings following the eruption. However, there is only ambiguous evidence for degassing of $CO_2$, synchronous with the reduced $\delta^{18}O$ values, which could have reduced the $^{14}C$ content in tree rings of 2001 and 2002. Our results illustrate that additional water available to the trees may have promoted photosynthesis and carbon fixation and also suggest photosynthetic response to pre-eruption $CO_2$ degassing as an additional possible explanation for local increases in NDVI before the 2002/2003 Mt. Etna flank eruption (Houlié et al., 2006). NDVI might be used as indicator of volcanic precursory activities.

The study of the impact of volcanic activity on past climates and the global carbon cycle is compromised by the paucity of annually resolved proxy records (Büntgen et al., 2020). New proxies that may provide more detailed information than those already used, e.g., ring width, are needed (e.g., Edwards et al., 2021). Tree-ring oxygen stable isotopes can be used as indicators of past volcanic eruptions.

## ACKNOWLEDGEMENTS

We dedicate this paper in memory of our colleague and friend, Sebastiano Cullotta (University of Palermo), who tragically passed away while this paper was in preparation. Sebastiano introduced us to the secrets of the magical woodlands growing among the silent lava flows on the Etna's slopes.

We thank our colleague Anne Verstege for her continuous support in the dendrochronological laboratory. Also we wish to thank Vincenzo Crimi (Corpo Forestale, Bronte) for his logistical support during our fieldwork on Mt. Etna, as well as the Corpo Forestale della Regione Siciliana (Distaccamento di Bronte, Catania, Italy) for their permission to take samples.

## CONFLICT OF INTEREST

The authors have no competing interests to declare.

## FUNDING INFORMATION

This work was supported by the Swiss National Science Foundation (205321_143479).

## DATA AVAILABILITY STATEMENT

The data that support the findings of this study are available from the corresponding author upon reasonable request.

## ORCID

*Paolo Cherubini*  https://orcid.org/0000-0002-9809-250X